\documentclass{ws-ijgmmp}

\begin{document}

\markboth{Micha\l{} Szcz\c{a}chor}
{Supersymmetric Holst action with matter coupling and parity violation}

%%%%%%%%%%%%%%%%%%%%% Publisher's Area please ignore %%%%%%%%%%%%%%%
%
\catchline{}{}{}{}{}
%
%%%%%%%%%%%%%%%%%%%%%%%%%%%%%%%%%%%%%%%%%%%%%%%%%%%%%%%%%%%%%%%%%%%%

\title{SUPERSYMMETRIC HOLST ACTION WITH MATTER COUPLING AND PARITY VIOLATION
}

\author{MICHA\L{} SZCZ\c{A}CHOR}%\footnote{
%Typeset names in 8 pt roman, uppercase. Use footnote to indicate 
%permanent address of author.}}

\address{Institute for Theoretical Physics, University of Wroc\l{}aw\\
Pl. Maksa Borna 9, Pl-50-204 Wroc\l{}aw, Poland\\
%University Department, University Name, Address\\
%City, State ZIP/Zone,Country\\%\,\footnote{State completely without 
%abbreviations, the affiliation and mailing address, including country. 
%Typeset in 8 pt italic.}\\ 
\email{misza@ift.uni.wroc.pl} }%\footnote{Typeset author's e-mail 
%address in 8pt italic}} }

%\author{SECOND AUTHOR}

%\address{Group, Laboratory, Address\\
%City, State ZIP/Zone, Country\\
%author\_id@domain\_name }

\def\bp{\partial\!\!\!/}
\def\bop[#1]{\overleftrightarrow{\partial}_{#1}}

\maketitle

%\begin{history}
%\received{(Day Month Year)}
%\revised{(Day Month Year)}
%\end{history}

\begin{abstract}
%The general construction of the Holst action is presented. Basing on the this construction the super symmetric $N=1,2,4,8$ Holst action is studied. 
%The minimal supersymmetric model $N=1$ was extended for matter coupled in the way preserving the supersymmetry. The obtained result shows that supersymmetry preserves 
%appearance of the Immirzi parameter in the equation of motion. This is in contradiction to the case of minimal/non-minimal coupling of non-supersymmetric
%Dirac spinors, where the Immirzi parameter is present in the equation of motion. The construction ensures that the theory is invariant under supersymmetry as well as gauge 
%transformation, but the extension breaks the parity of the action.  
A general construction of the Holst action is discussed. Based on this, the $N=1,2,4,8$ supergravities and $N=1$ supergravities with matter coupling are presented.
It is shown that in all this cases the Immirzi parameter does not influence the field equations. The construction ensures that the theory is invariant under supersymmetry as well as gauge 
transformations, but the Holst extension breaks parity.  
\end{abstract}

\keywords{SUGRA; Holst action; matter coupling; parity violation.}

\section{Introduction}
\numberwithin{equation}{section}
The Holst action is an action used as a starting point to quantize gravity in theories employing real $SU(2)$ connection. In fact, Holst term is much
older then Holst action formulation or even the Loop Quantum Gravity (LQG). It was first presented in \cite{Hojman:1980kv} as a possible extension of the Hilbert-Einstein 
action invariant under gauge symmetries (e.g. Lorentz and diffeomorphisms symmetries) but breaking parity whenever the torsion is present. It turns out that the extension of the 
Lagrangian density $R$ by the term $\epsilon^{\mu\nu\rho\sigma}R_{\mu\nu\rho\sigma}$ which is proposed in this article is the only possible extension linear in R in the Einstein-Cartan framework.
It was suggested at the end of \cite{Hojman:1980kv} that it would be interesting to extend the results to the supersymmetric case. Such extension is presented at this work. 

%This work also shows how the symmetry preserves a physical behavior. The coupling of matter is one of the issues needed to approve LQG as a promising approach to quantized gravity. 
 The minimal/non-minimal coupling  of the matter fields with spin-$1/2$ was already studied in the Holst action framework by several
authors \cite{Perez:2005pm,Freidel:2005sn}. It was shown that the term with Immirzi parameter becomes dynamical i.e. it appears in the equation of motion (e.o.m.) and the parity is violated as expected. However, in \cite{Mercuri:2006um} it has been shown that there is a coupling of fermions to the Holst action which does not affect the equations of motion\footnote{In the Ref. \cite{Mercuri:2006um} it was shown that the Holst action can be rewritten as $\int d^4x \,(NY + \partial_\mu J^\mu)$, where $NY$ is the Nieh-Yan topological invariant and $J^\mu$ is a current of matter field. This property is more general and can be extended for derivations contained at this work. This turns out to be equivalent to the argument that if we take into account the terms quadratic in contorsion in the dynamical part of the action as well as in the Holst term (see \ref{appp}), the Immirzi parameter is not dynamical. }
.
 In spite of the interest in coupling fermions to the theory, the 
supersymmetric Holst action was studied only in \cite{Kaul:2007gz} as an extension of the Holst bosonic Lagrangian or in the \cite{Durka:2009pf} where the supersymmetric 
Holst action was obtained as a result of super--BF theory with constrains\footnote{super--BF theory of the  MacDowell--Mansouri type}. 
In both cases it turned out that the Holst term does not influence the e.o.m. As this work
shows, these properties are general features of supersymmetric theories. 

To clarify some misunderstanding in the role of the Holst term and Immirzi parameter, this paper will first recall the way of constructing Holst term, then will shortly present 
the role of the Immirzi
parameter. The last part of this section is devoted to rudiments of general theory concerns the construction of the general parity violation term, based on the torsion equation. In the second part of this work the 
construction for higher supersymmetry will be presented and the super-Holst action $N=1$ will be coupled, using the Noether method, to scalar-spinor-$1/2$, Maxwell and Yang--Mills multiplet.% basis on 
%the general theory.  

\section{Appearance of Immirzi parameter in General Relativity}

The Immirzi parameter was introduced by Barbero \cite{Barbero:1994ap} and Immirzi\footnote{Immirzi parameter is also named as Barbero--Immirzi parameter to honor both authors.}  \cite{Immirzi:1996di}
as a parameter which allows to introduce real $SU(2)$ connection\footnote{The group $SU(2)$ is double covering of the group $SO(3,1)$.} instead of complex $SL(2,C)$ Ashtekar connection \cite{Ashtekar:1986yd} in canonical quantization of general relativity.   

The construction follows from noticing that if one partially fixes the $SO(3,1)$ gauge freedom by "time gauge``, what is left is the invariance under local $SO(3)$, with frame space--like 
field $e^{i\mu}$($\mu,\nu,\dots$ are space indexes, and $i,j,...=1,2,3$ internal indices), and 3--metric constructed from frame field $q^{\mu\nu}=e^{i\mu}e^{i\nu}$. Then one can 
construct a densitized triad $E^{i\mu}=\frac{1}{2}\epsilon^{\mu\nu\rho}\epsilon_{ijk}e^j_\nu e^k_\rho=\sqrt{q}e^{i\mu}$. There are two connection present in this construction, a $SO(3)$
Levi--Civita connection $\Gamma^i_\mu=\frac{1}{2}q_\mu\,^\nu\epsilon_{jik}\omega^{jk}_\nu$ and the connection $K^i_\mu=q_\mu\,^\nu\omega^{0i}_\nu=e^{i\nu}K_{\mu\nu}$ (the $\omega^{IJ}_\mu$  is a spin connection where $I,J,\dots=0,1,2,3$) conjugated to $E^{i\mu}$ i.e.
\begin{align}
 \{E^{i\mu}(x), K^{j}_\nu(y) \}=\delta^{ij}\delta^\mu_\nu\delta^3(x,y)\, .
\end{align}
Barbero and Immirzi pointed out \cite{Barbero:1994ap, Immirzi:1996di},  that there exist a cannonical transformation of variables 
\begin{align}
&&& E^{i\mu}\to E^{i\mu}&\\
&&&A^i_\mu\to \Gamma^i_\mu+\gamma K^i_\mu& \label{Bar}
\end{align}
which results in a single non vanishing Poissone brackets
\begin{align}
 \{E^{i\mu}(x), A^j_\nu(y)\}=\gamma\, \delta^{ij}\delta^\mu_\nu\delta^3(x,y)\, .
\end{align}
The parameter $\gamma$ is called Immirzi parameter and its presence ensures that the theory has a Lorentzian signature as long as $\gamma\neq 1$ (which is Euclidean case). If one sets $\gamma=\pm i$
then one reconstructs the original Ashtekar self/anti-self dual variables \cite{Ashtekar:1986yd}. For $\gamma\in \mathbb{R}-\{0\}$ the variables are real. 

The Immirzi parameter also appears in a more general context in \cite{Wise:2009fu}, where it was shown that by using the most general inner product on Lie algebra $so(3,1)$
\begin{align}
 <X,Y>=tr(X(c_0+\star\, c_1)Y)\, ,
\end{align}
and identifying $(c_0,c_1)=(1,\frac{1}{\gamma})$, one gets a theory, that includes the Immirzi parameter\footnote{More precisely the Holst action (which is presented in section \ref{HolstA}) with boundary terms.}. 
 
\section{Holst action \label{HolstA}}

An action which leads to Barbero formulation in Hamiltonian framework \cite{Barbero:1994ap} is the action proposed in \cite{Holst:1995pc} \footnote{The $\alpha$ was introduced for simplification as it is a numerical factor.}
\begin{align}
 S=\frac{1}{\alpha G} \int e e^\mu_I e^\nu_J (R_{\mu\nu}^{IJ}-\beta \star R_{\mu\nu}^{IJ})\, .\label{holst}
\end{align}
This action was proposed earlier \cite{Hojman:1980kv} in the form 
\begin{align}
 S=\int d^4x\, \frac{1}{16 \pi G}\sqrt{-g}R+\frac{1}{16\pi G_P}\epsilon^{\mu\nu\rho\sigma}R_{\mu\nu\rho\sigma}\, , 
\end{align}
 where $G_P\sim \gamma$ governs the parity violating interaction. 

Observe that the action (\ref{holst}) is based on the Barbero connection.
One can take a variation of action
\begin{align}
 \delta_\omega S=\frac{1}{\alpha G}\int e e^\mu_I e^\nu_\nu \ \delta_\omega( R_{\mu\nu}^{IJ}-\frac{\beta}{2}\epsilon^{IJ}\,_{KL}  R_{\mu\nu}^{KL})=\frac{2}{\alpha G}\int \delta_\omega B^{IJ}_\nu D_\mu^\omega(e\,e^\mu_Ie^\nu_J) 
\end{align}
where 
\begin{align}
\label{conss} \delta_\omega B_\nu^{IJ}=\delta_\omega \omega_\nu^{IJ}-\beta\star\delta_\omega \omega_\nu^{IJ}\,.
\end{align}
In the case where $\beta=\pm i$ one can recognizes the Ashtekar connection. 
If one uses the space-time decomposition of connection, i.e.
\begin{align}
&&& \omega_{\mu ij}=-\epsilon_{ijk}\Gamma^k_\mu=-\epsilon_{ijk}(-\frac{1}{2}\epsilon^{klm}e_{\nu l}\nabla_\mu e^\nu_k)&\\
&&&\omega_{\mu i0}=e^\nu_i\nabla_\mu e_{\nu 0}=K_{\mu i}\, &
\end{align}
and inserts it to the action, 
then it is easy to see that action depends exactly on the Barbero connection 
\begin{align}
 A_{\mu i}=\Gamma_{\mu i}+\frac{1}{\beta} K_{\mu i}\, ,
\end{align}
with the identification $\frac{1}{\beta}=\gamma$.     

\section{General construction of the Holst term\label{se2}}

The construction of connection (\ref{conss}) suggests how the Holst term is constructed. It is built from terms belonging to the torsion equation i.e. from e.o.m. calculated under variation
$\delta_\omega$, multiplied by $\frac{1}{\gamma}\,\star$. The torsion equation has the following general structure 
\begin{align}
T_{[\mu\nu]}\,^J=D_{[ \mu} e_{\nu ]}^J=K_{[ \mu}\,^{IJ}e_{\nu ] J}=-2K_{[\mu\nu]}\,^I\,.\label{tortor}
\end{align}
The tensor $K_\mu\,^{IJ}$ is called contorsion and is  defined as a difference between torsion and torsion-free connection
\begin{align}
 \Gamma^\lambda_{\mu\nu}=\tilde{\Gamma}^\lambda_{\mu\nu}-K_{\mu\nu}\,^\lambda\,.\label{tradd}
\end{align}
In terms of spin connection it can be rewritten as
\begin{align}
 \omega^{IJ}_\mu=\omega^{IJ}_\mu+K^{IJ}_\mu\,.\label{omee}
\end{align}
 The contorsion $K_{\mu\nu}\,^\lambda$ can be obtained from
\begin{align}
 e^\rho_IT^I\,_{\mu\nu}(e,\omega(e)+K)=-2 K_{[\mu\nu]}\,^\rho
\end{align}
and can be expressed in term of torsion as
\begin{align}
 K_{\mu\nu}\,^\rho=\frac{1}{2}(T_{\mu\nu}\,^\rho-T^\rho\,_{\mu\nu}+T_{\nu\mu}\,^\rho)\, .
\end{align}
%\begin{theorem}\label{t1}

The Holst term can be constructed from torsion and contorsion tensor by the following contractions
\begin{align}
&&\nonumber S^{Holst}&=\frac{1}{G}\int\frac{1}{\gamma}\epsilon^{\mu\nu\rho\sigma} (T_{\mu\nu}\,^I+K_{\mu\nu}\,^I) (T_{\rho\sigma I}+K_{\rho\sigma I})&\label{hoholst}\\
&&& =\frac{1}{G}\int\frac{2}{\gamma}(T_{\mu\nu}\,^I+K_{\mu\nu}\,^I)\star (T^{\mu\nu}\,_I+K^{\mu\nu}\,_I)\,.&
\end{align}
%\end{theorem}
%\begin{proof}
It arises from the (\ref{conss}) that a variation with respect to $\omega$ is
%\begin{align}
% \delta_\omega S=\frac{2}{\alpha G}\int \epsilon^{\mu\nu\rho\sigma}\epsilon_{IJKL}\delta\omega_{\rho\sigma I}\,(1-\frac{1}{\gamma}\star)(T_{\mu\nu}\,^I-K_{\mu\nu}\,^I)\,.
%\end{align}
\begin{align}\label{rty}
 \delta_\omega S=\frac{4}{\alpha G}\int \epsilon^{\mu\nu\rho\sigma}\epsilon_{IJKL}\delta\omega_{\sigma}\,^{IJ}\Big(\,(T_{\rho\mu}^{K}e^L_\nu+K_{\rho\mu}^{K}e^L_\nu)-\frac{1}{\gamma}\star(T_{\rho\mu}^{K}e^L_\nu+K_{\rho\mu}^{K}e^L_\nu)\Big)\,.
\end{align}
%The first part of above calculation gives the result for which the equation of torsion (\ref{tortor}) is a solution. 
%The second part of above calculation 
If one restricts oneself to a theory linear in $R$ and without any second covariant derivatives, the most general object of 3 indices  (2 space-time and 1 internal) 
which can be contracted via $\epsilon^{\mu\nu\rho\sigma}$ and gives the second part of (\ref{rty})
\begin{align}
% \delta_\omega S=
-\frac{4}{\alpha G}\int \epsilon^{\mu\nu\rho\sigma}\Big(\delta\omega_{\sigma\nu I}\,\frac{2}{\gamma}\,(T_{\rho\mu}^{I}+K_{\rho\mu}^{I})\Big)\,
\end{align} 
is exactly (\ref{hoholst}). 
%$T_{\rho\mu}^{I}+K_{\rho\mu}^{I}$. 
\\
%\end{proof}

There are some comments on this construction. First, if matter fields are not included to Lagrangian, there is no torsion \cite{Hojman:1980kv}. In the other words the torsion
\begin{align}
 T_{\mu\nu}\,^\lambda=\Gamma_{[\mu\nu]}^\lambda
\end{align}
is antisymmetric in lower indices and there is no contorsion term $K^{\lambda\mu\nu}=0$. 
To get dynamical torsion, due to the fact that torsion couples to spin and gravity field
has spin-2, one can express contorsion field in a term of scalar field 
\begin{align}
 K_{\alpha\beta\gamma}=\phi_{,\beta}g_{\alpha\gamma}-\phi_{,\gamma}g_{\alpha\beta}
\end{align}
and ensure that this equation will be satisfied by adding it to the action via Lagrange multiplier. However this procedure does not break parity symmetry \cite{Hojman:1980kv}. 
In the case of matter field the term (\ref{hoholst}) bieng a pseudoscalar breaks parity. Second
comment is that there exist much more terms quadratic in contorsion which can be added to the action (see \cite{Hojman:1980kv, Benedetti:2011nd}), even the terms added in
nondynamical manner, which results in the dynamical torsion i.e.  terms with second derivatives of $K$ or equivalently, terms quadratic on the derivatives of $K$ \cite{Hojman:1980kv, Neville:1979rb}.    

\section{Supersymmetric $N=1,2,4,8$ Holst action}

Based on the construction (\ref{hoholst}), the supersymmetric extension\footnote{The action and torsion equation for case $N=8$ can be found in \cite{de Wit:1982ig}} of the Holst action can be easily constructed and classified\footnote{The terms agree with published in \cite{Durka:2009pf} for SUGRA $N=1$ and in \cite{Kaul:2007gz} for SUGRA $N=1,2,4$.}. In the case of supergravity (SUGRA)
expression in (\ref{tortor}) is called ''super torsion``\footnote{From this section $i=0,\dots,3$ are internal indices, $\mu,\nu=0,\dots,3$ are space-time indices and $I,J,K$ numbers the different supersymmetric charges.}.
\begin{table}[ht]
\tbl{Super torsion equation and Holst term for different SUGRA theories.}
{\begin{tabular}{@{}ccl@{}} \toprule
Supergravity & Super torsion & Holst term  \\
$N$ &  \\ \colrule
1 & $ T_{\nu\rho}{}^{a}=\bar\psi_\nu\gamma^a\psi_\rho $ & $ e\left[e^{\mu}_a e^{\nu}_b~{\tilde R}_{\mu\nu}^{~~~ab} - \,\frac{1}{e}\epsilon^{\mu\nu\rho\sigma}\bar{\psi}_\mu\, \gamma_{a}\, \psi_\nu \, D^\omega_\rho e_\sigma{}^a\right]\,
$   \\ \colrule
2 & $ T_{\mu\nu}^{~~~a}(\psi)=\left( {\bar \psi}^I_\mu \gamma^a_{~}
\psi^{~}_{I\nu} ~+~ {\bar \psi}^{~}_{I\mu} \gamma^a_{~} \psi^I_\nu \right)$ & $ e\left[ e^{\mu}_a e^{\nu}_b~{\tilde R}_{\mu\nu}^{~~~ab}
~-~\frac{1}{2e} ~\epsilon^{\mu\nu\alpha\beta} {\bar \psi}^I_\mu \psi^J_\nu  
~{\bar \psi}^{~}_{I\alpha} \psi^{~}_{J\beta}~ \right.$ \\
&& $\left. -~\frac{1}{e}\epsilon^{\mu\nu\alpha\beta}  \left( 
{\bar \psi}^I_\mu \gamma^{~}_\nu D^{~}_\alpha\psi^{~}_{I\beta}~ +~ {\bar \psi}^{~}_{I\mu} 
\gamma^{~}_\nu D^{~}_\alpha\psi^I_\beta \right)\right]   $   \\ \colrule
4 & $T_{\mu\nu}^{~~a} = {\bar \psi}^I_{[\mu} \gamma^a_{} \psi^{}_{\nu ]I}
+ {\frac 1 {e}} e^{a \alpha}_{} \epsilon_{\mu\nu\alpha\beta}^{} 
{\bar \lambda}^{}_I \gamma^\beta_{} \lambda^I_{}$ & 
$e\left[ ~ e^{\mu}_a e^{\nu}_b
~{\tilde R}_{\mu\nu}^{~~~a b} - {\frac 1 {e}}~ \epsilon^{\mu\nu\alpha\beta}_{~}
\left({\bar \psi}^I_\mu \gamma^{~}_\nu { D}^{~}_\alpha \psi_{I\beta} + {\bar \psi}^{~}_{I\mu} \gamma^{~}_\nu { D}^{~}_\alpha  \psi^I_\beta \right) \right.$  \\
 &  & $ -  ~ \left({\bar \lambda}^{~}_I \gamma^\mu_{~} 
{ D}_\mu^{~} (\omega) \lambda^I_{~}
-{\bar \lambda}^I_{~} \gamma^\mu_{~} { D}_\mu^{~} (\omega) \lambda^{~}_I \right) $ \\
&& $ \left. -{\frac 1 {2e}} \epsilon^{\mu\nu\alpha\beta}_{~} ~{\bar \psi}^I_\mu \psi^J_\nu{\bar \psi}^{~}_{I\alpha} \psi^{~}_{J\beta} 
 -{\frac 1 {2e}} ~\epsilon^{\mu\nu\alpha\beta}_{~}
~{\bar \lambda}^I \gamma_\mu \psi^J_\nu ~{\bar \lambda}_I \gamma_\alpha \psi_{J\beta} \right]$ \\ \colrule
8 & $T_{\mu\nu}^{~~a} = \bar{\psi}^I_{[\mu}\gamma^a\psi_{\nu] I}+\frac{1}{12}\epsilon_{\mu\nu}\,^a\,_b \bar{\chi}^{IJK}\gamma^b\chi_{IJK}$ & $
e\left[ ~ e^{\mu}_{a} e^{\nu}_{b}
~{\tilde R}_{\mu\nu}^{~~~a b} - {\frac 1 {e}}~ \epsilon^{\mu\nu\alpha\beta}_{~}
~\left({\bar \psi}^I_\mu \gamma^{~}_\nu { D}^{~}_\alpha \psi^{~}_{I\beta}
+ {\bar \psi}^{~}_{I\mu} \gamma^{~}_\nu { D}^{~}_\alpha  \psi^I_\beta \right) \right.$\\
&& $\left. -  {\frac  1 {12}}\left({\bar \chi}^{~}_{IJK} \gamma^\mu_{~} 
{ D}_\mu^{~}  \chi^{IJK}_{~}
-{\bar \chi}^{IJK}_{~} \gamma^\mu_{~} { D}_\mu^{~} \chi^{~}_{IJK} \right)\right.$ \\ 
&&  $   
 -~ {\frac 1 {24e}} ~\epsilon^{\mu\nu\alpha\beta}_{~}
~{\bar \chi}^{IJK} \gamma_\mu \psi^L_\nu ~{\bar \chi}_{IJK} \gamma_\alpha \psi_{L\beta} $ \\ 
&& $  +~\frac{1}{144e}\epsilon_{\mu\nu}\,^a\,_c\epsilon_{\rho\sigma a b}(\bar{\chi}^{IJK}\gamma^c\chi_{IJK})(\bar{\chi}^{IJK}\gamma^b\chi_{IJK})$\\ 
&& $\left.  -{\frac 1 {2e}}~ \epsilon^{\mu\nu\alpha\beta}_{~} ~{\bar \psi}^I_\mu \psi^J_\nu 
~{\bar \psi}^{~}_{I\alpha} \psi^{~}_{J\beta} \right] $ \\ \botrule
\end{tabular}}
\end{table}

\section{Coupling matter to SUGRA $N=1$}

Matter can be coupled in the supersymmetric manner i.e. in the way which ensure that the Lagrangian will be invariant under supersymmetric, Lorentz and diffeomorphisms transformations.
The method of coupling is called the Noether method and details can be found in \cite{Ferrara:1976kg, VanNieuwenhuizen:1979hma, Ferrara:1976ni}.
 
The brief review of the method \cite{Ferrara:1976ni,VanNieuwenhuizen:1979hma} for the scalar-spin-$1/2$ multiplet based on consideration of Wess-Zumino Lagrangian which is scalar-spinor-$1/2$ theory with global supersymmetric invariance 
\begin{equation}
\mathcal{L}=-\frac{1}{2}eg^{\mu\nu}(\partial_\mu A\partial_\nu A + \partial_\mu B\partial_\nu B)- \frac{1}{2}e\bar{\chi} \gamma^\mu D_\mu \chi\, ,
\end{equation}
where $A$ is scalar, $B$ is pseudoscalar and $\chi$ is spin-$\frac{1}{2}$ spinor.
If one replaces a global supersymmetric invariance with local, the coupling term, accurate to second order in coupling constant $\kappa$, can be expressed as
\begin{eqnarray}
\mathcal{L}^{\kappa+\kappa^2}&=&\frac{1}{2} e\kappa\bar{\psi}_\mu(\bp A+i\gamma_5\bp B)\gamma^\mu\chi-\frac{1}{16}e\kappa^2(\bar{\chi}\chi)^2\\
&=&\frac{1}{32}(\bar{\chi}\gamma_5\gamma_\tau\chi)[\epsilon^{\mu ab\tau}\bar{\psi}_\mu\gamma_a \psi_b-2e(\bar{\psi}_\alpha\gamma_5\gamma^\tau\psi^\alpha)]\\
&=&\frac{1}{8}i\kappa^2(A \bop[\beta] B)[e\bar{\chi}\gamma_5\gamma^\beta\chi-\epsilon^{\alpha\beta\mu\tau}\bar{\psi}_\alpha\gamma_\tau\psi_\mu]
\end{eqnarray}
and the supersymmetric transformation, accurate to first order in $\kappa$, are
\begin{equation}
\delta\chi=\bp(A+i\gamma_5B)\epsilon-\frac{1}{2}\kappa(\bar{\psi}_\rho)\gamma^\rho\epsilon-\frac{1}{2}\kappa(\bar{\psi}_\rho\gamma_5\chi)\gamma_5\gamma^\rho\epsilon
\end{equation}
\begin{equation}
\delta \psi_\mu=\frac{2}{\kappa}D_\mu\epsilon+\frac{i\kappa}{2}\gamma_5\epsilon(A\bop[\mu] B)+\frac{\kappa}{4}(\chi\gamma_5\gamma^\rho\chi)(\sigma_{\mu\rho}\gamma_5\epsilon)
\end{equation}
\begin{equation}
\delta A=\bar{\epsilon}\chi\qquad \delta B=i\bar{\epsilon}\gamma_5\chi\qquad \delta e_\mu \,^i=i\kappa\bar{\epsilon}\gamma^i\psi_\mu 
\end{equation}
\begin{equation}
 \delta \omega^{ij}=-\kappa\bar{\epsilon}\gamma^{ij}\psi_\mu\ . 
\end{equation}
The total Lagrangian of supersymmetric gravity coupled to scalar multiplet is
\begin{eqnarray}
\label{l1}
\nonumber \mathcal{L}&=&\mathcal{L}^{SUGRA}-\frac{1}{2}eg^{\mu\nu}(\partial_\mu A\partial_\nu A + \partial_\mu B\partial_\nu B)- \frac{1}{2}e\bar{\chi} \gamma^\mu D_\mu \chi\\
\nonumber &+&\frac{1}{2}e\kappa\bar{\psi}_\mu(\bp A+i\gamma_5\bp B)\gamma^\mu\chi-\frac{1}{16}e\kappa^2(\bar{\chi}\chi)^2\\
\nonumber &+&\frac{1}{32}(\bar{\chi}\gamma_5\gamma_\tau\chi)[\epsilon^{\mu ab\tau}\bar{\psi}_\mu\gamma_a \psi_b-2e(\bar{\psi}_\alpha\gamma_5\gamma^\tau\psi^\alpha)]\\
&+&\frac{1}{8}i\kappa^2(A \bop[\beta] B)[e\bar{\chi}\gamma_5\gamma^\beta\chi-\epsilon^{\alpha\beta\mu\tau}\bar{\psi}_\alpha\gamma_\tau\psi_\mu]\, .
\end{eqnarray}
The equation of torsion can be calculated from (\ref{l1}) as
\begin{eqnarray}
\label{tt2}T_{\nu\rho i}&=&D^\omega_{[\nu}e^{\null}_{\rho]i}=\bar{\psi}_\nu\gamma_i\psi_\rho+\frac{1}{e}e^\mu_i\epsilon_{\nu\rho\mu\sigma}\bar{\chi}\gamma^\sigma\chi\,.
\end{eqnarray}
Using the construction (\ref{hoholst}) the Holst term reads
\begin{eqnarray}
\label{holhol}
\nonumber \mathcal{L}^{Holst}&=&\frac{1}{\gamma}\epsilon^{\mu\nu\rho\sigma}\Big(4D_\nu e_{\rho i}D_\mu e_{\sigma}^i- \bar{\psi}_\mu\gamma_i\psi_\nu D_\rho e_\sigma^i-\frac{1}{e}e^\alpha_i\epsilon_{\mu\nu\alpha\beta}\bar{\chi}\gamma^\beta\chi D_\rho e_{\sigma }^i+\frac{1}{4}\bar{\psi}_\mu\gamma_i\psi_\nu \bar{\psi}_\rho\gamma^i\psi_\sigma\\
 &&+\frac{1}{8e}e^\alpha_i\epsilon_{\mu\nu\alpha\beta}\bar{\chi}\gamma^\beta\chi\bar{\psi}_\mu\gamma^i\psi_\nu+ \frac{1}{4e^2}e^\alpha_i\epsilon_{\mu\nu\alpha\beta}\bar{\chi}\gamma^\beta\chi e^{\delta i}\epsilon_{\rho\sigma\delta\xi}\bar{\chi}\gamma^\xi\chi \Big).
\end{eqnarray}
%\begin{eqnarray}
%\label{holhol}
%\nonumber L^{Holst}&=&\epsilon^{\mu\nu\rho\sigma}\Big(\frac{4}{\gamma G}D_\nu e_{\rho i}D_\mu e_{\sigma j}-\frac{i}{2\gamma} \bar{\psi}_\mu\gamma_i\psi_\nu D_\rho e_\sigma\,^i~~~~~~~~~~\\
%\nonumber &&-\frac{l}{4G\gamma}e^{\sigma l}\Big(\frac{\gamma^2}{1+\gamma^2}\Big)\epsilon_{\mu\nu\rho\sigma}(\frac{1}{\gamma}\bar{\chi}\gamma^\mu\sigma_{kl}\chi+\epsilon_{ijkl}\bar{\chi}\gamma^\mu\sigma^{ij}\chi)
%D_\mu e_{\sigma j}\\
%\nonumber &&+\frac{i}{2\gamma}\frac{1}{4}e^{\sigma l}\Big(\frac{\gamma^2}{1+\gamma^2}\Big)\epsilon_{\mu\nu\rho\sigma}(\frac{1}{\gamma}\bar{\chi}\gamma^\mu\sigma_{kl}\chi+\epsilon_{ijkl}\bar{\chi}\gamma^\mu\sigma^{ij}\chi)\bar{\psi}_\rho\gamma^i\psi_\nu\\
%\nonumber &&-\frac{l}{2G\gamma}\frac{1}{16}e^{\sigma l}\Big(\frac{\gamma^2}{1+\gamma^2}\Big)e^{\delta n}\Big(\frac{\gamma^2}{1+\gamma^2}\Big)\epsilon_{\mu\nu\rho\sigma}(\frac{1}{\gamma}\bar{\chi}\gamma^\mu\sigma_{kl}\chi+\epsilon_{ijkl}\bar{\chi}\gamma^\mu\sigma^{ij}\chi)~~~~~~~~\\
%&&\cdot\,\epsilon_{\chi\nu\rho\delta}(\frac{1}{\gamma}\bar{\chi}\gamma^{\chi}\sigma_{kn}\chi+\epsilon_{ijkn}\bar{\chi}\gamma^\mu\sigma^{ij}\chi)\Big)\ .~~~~~~~
%\end{eqnarray}

One can couple the Maxwell field using exactly the same procedure \cite{Ferrara:1976ni}. The action for coupled abelian field in supersymmetric way is
\begin{eqnarray}
\nonumber \mathcal{L}&=&\mathcal{L}^{SUGRA}+\frac{1}{4}eg^{\mu\rho}g^{\nu\sigma}F_{\mu\nu}F_{\rho\sigma}-\frac{1}{2}e\bar{\lambda}\gamma^\mu {D_\mu} \lambda+\frac{1}{4}e\kappa\bar{\psi}_\mu\gamma^\alpha\gamma^\beta\gamma^\mu\lambda F_{\alpha\beta}\\
&&\nonumber+\frac{1}{8}e\kappa^2\Big(-(\bar{\psi}\cdot\psi)(\bar{\lambda}\lambda)+(\bar{\psi}_\alpha\gamma^5\psi^\alpha)(\bar{\lambda}\gamma^5\lambda)\\
&&\nonumber+\frac{1}{4}(\bar{\psi}\cdot\gamma\gamma\cdot\psi)(\bar{\lambda}\lambda)+\frac{1}{4}(\bar{\psi}\cdot\gamma\gamma^5\gamma\cdot\psi)(\bar{\lambda}\gamma^5\lambda)\\
&&\nonumber-\frac{1}{2}(\bar{\psi}\cdot\gamma\gamma^5\cdot\psi_\alpha)(\bar{\lambda}\gamma^5\gamma^\alpha\lambda)+\frac{1}{4}(\bar{\psi}_\alpha\cdot\gamma_5\gamma_\rho\cdot\psi^\alpha)(\bar{\lambda}\gamma_5\gamma^\rho\lambda)\\
&&+\frac{3}{2}(\bar{\lambda}\lambda)(\bar{\lambda}\lambda)\Big)\, .
\end{eqnarray}
One can recognize the two connections, a spin connection and an abelian connection, but the impact to super torsion equation is given only by a spin connection and it looks exactly
the same as (\ref{tt2}).  

If one considers a Yang--Mills multipled given by the Lagrangian
\begin{eqnarray}
\nonumber \mathcal{L}&=&\mathcal{L}^{SUGRA}+\frac{1}{4}eg^{\mu\rho}g^{\nu\sigma}F_{\mu\nu}F_{\rho\sigma}-\frac{1}{2}e\bar{\lambda}\gamma^\mu {D_\mu} \lambda+\frac{1}{4}e\kappa\bar{\psi}_\mu\gamma^\alpha\gamma^\beta\gamma^\mu\lambda F_{\alpha\beta}\\
&&\nonumber+\frac{1}{8}e\kappa^2\Big(-(\bar{\psi}\cdot\psi)(\bar{\lambda}\lambda)+(\bar{\psi}_\alpha\gamma^5\psi^\alpha)(\bar{\lambda}\gamma^5\lambda)\\
&&\nonumber+\frac{1}{4}(\bar{\psi}\cdot\gamma\gamma\cdot\psi)(\bar{\lambda}\lambda)+\frac{1}{4}(\bar{\psi}\cdot\gamma\gamma^5\gamma\cdot\psi)(\bar{\lambda}\gamma^5\lambda)\\
&&\nonumber-\frac{1}{2}(\bar{\psi}\cdot\gamma\gamma^5\cdot\psi_\alpha)(\bar{\lambda}\gamma^5\gamma^\alpha\lambda)+\frac{1}{4}(\bar{\psi}_\alpha\cdot\gamma_5\gamma_\rho\cdot\psi^\alpha)(\bar{\lambda}\gamma_5\gamma^\rho\lambda)\Big)\\
&&+\frac{3}{64}e\kappa^2(Tr(\bar{\lambda}\gamma_5\gamma_\rho\lambda))^2\, , 
\end{eqnarray}
than immediately will see, that using non-abelian matter fields instead of abelian do not influence to torsion equation (\ref{tt2}). Therefore, it is a general observation that Holst term in the scalar-spin-$1/2$, Maxwell and Yang--Mills theory has the same form of (\ref{holhol}). 
\section{General Theory for SUGRA}
It is possible to formulate general theory, which is valid for any supersymmetric matter couple to SUGRA N=1 and for any higher SUGRA (N=1,2,4,8).  
\begin{theorem}\label{t2}
 The Holst term constructed in SUGRA $N=1,2,4,8$ and SUGRA $N=1$ coupled to matter is invariant under supersymmetric, Lorentz and diffeomorphisms transformation and does not
influence the field equations.  
\end{theorem}
\newpage
\begin{proof}
The torsion equation has a general form 
\begin{align}\label{heh}
 T_{\mu\nu}^i+K_{1\mu\nu}^i+\cdots+K_{N\mu\nu}^i=0\, ,
\end{align}
where $N$ is a number of different matter sources present in a theory. Using the construction (\ref{hoholst}) the Holst term reads
 \begin{eqnarray}
\nonumber \mathcal{L}&=&\epsilon^{\mu\nu\rho\sigma}(\frac{1}{2}T_{\mu\nu}^iT_{i\rho\sigma}+T_{\mu\nu}^iK_{1i\rho\sigma}+\cdots+T_{\mu\nu}^iK_{Ni\rho\sigma}\\
\nonumber&+&\frac{1}{2}K_{1\mu\nu}^iK_{1i\rho\sigma}+\cdots+K_{1\mu\nu}^iK_{Ni\rho\sigma}\\
&+&\cdots+\frac{1}{2}K_{N\mu\nu}^iK_{Ni\rho\sigma})\ .
\end{eqnarray}
One can notices that any arbitrary infinitesimal variation of this Lagrangian
 \begin{eqnarray}
\nonumber\delta \mathcal{L}&=&\epsilon^{\mu\nu\rho\sigma}(\delta T_{1\mu\nu}^iT_{1i\rho\sigma}+2\delta T_{1\mu\nu}^iT_{2i\rho\sigma}+\cdots+2\delta T_{1\mu\nu}^iT_{Ni\rho\sigma}\\
\nonumber&+&\delta T_{2\mu\nu}^iT_{2i\rho\sigma}+\cdots+2\delta T_{2\mu\nu}^iT_{Ni\rho\sigma}\\
&+&\delta T_{N\mu\nu}^iT_{Ni\rho\sigma})\, 
\end{eqnarray}
vanishes according to torsion equation (\ref{heh}).
 
\end{proof}

\section{Conclusion}

One can realize that in the supersymmetry theories the terms depending on Immirzi parameter are not dynamical. This result is much different from many works on coupling spinor fileds to gravity \cite{Perez:2005pm,Freidel:2005sn}. 
Therefore, the understanding of the difference is very instructive. Usually, the strategy of coupling matter field to Holst action is based on adding the matter Lagrangian with 
some coupling constants. The significant is that the matter field does not contribute the Holst action. If one calculates the torsion equation then immediately realizes that it has some impact to torsion--free connection i.e. contorsion,
and contorsion depend on the Immirzi parameter. Solution of this equation gives the new connection (\ref{omee}), which contribute to action by some additional 
terms (see (\ref{a3}) in the Appendix) e.g. $K_{\mu\nu\rho}K^{\mu\nu\rho}$, which appear in e.o.m. Strategy presented in this work assumes that the Holst term is constructed
according to construction (\ref{hoholst}), which ensures that Holst term will be supersymmetric (using Theorem \ref{t2}), as the theory without Holst term has to be supersymmetric by
itself. Then if super torsion equation is calculated, contorsion tensor will not depend on Immirzi parameter anymore. Therefore, the action with additional terms (\ref{a2}, \ref{a3}) 
also does not depend on Immirzi parameter. Thus one can see that the assumption of supersymmetry does not allow appearance of the Immirzi parameter in e.o.m. 

However, considering the additional term \cite{Hojman:1980kv, Benedetti:2011nd, Neville:1979rb} in the super--Holst action is very interesting according to the interpretation of
Immirzi parameter as a regulator of the quantum fluctuations of the vanishing torsion condition. Author leaves this consideration for the future work.

\section*{Acknowledgments}

Author would like to thank  Wolfgang M. Wieland for reference \cite{Hojman:1980kv} and helpful discussion, Simone Mercuri for discussion and indication of the many interesting problems, Jerzy Kowalski--Gilkman for inspiration, discussion and comments on the
manuscript, and Tomasz Bia\l{}ach for some remarks. This work was partially supported by National Science Center grant No. 2011/01/N/ST2/00415.

\appendix

\section{The expanding tensor $R_{\mu\nu\rho\sigma}$ under full connection\label{appp}}
The construction (\ref{hoholst}) can be done alternatively \cite{Benedetti:2011nd} by calculating curvature using the (\ref{omee})
\begin{align}
\label{appi} e_{\mu K}e_{\nu L}R^{KL}_{\rho\sigma}(\omega(e)+K)=R_{\mu\nu\rho\sigma}(e)+2\nabla_{[\rho}K_{\sigma]\mu\nu}+K_{\rho\mu\lambda}K_\sigma\,^\lambda\,_\nu-K_{\sigma\mu\lambda}K_\rho\,^\lambda\,_\nu
\end{align}
and finding the dimension-two invariant of (\ref{appi}) under two possible contractions
\begin{align}
&\label{a2}\epsilon^{\mu\nu\rho\sigma} e_{\mu K}e_{\nu L}R^{KL}_{\rho\sigma}=(d^4x)K_{\mu\nu}\,^\lambda K_{\rho\sigma\lambda}\epsilon^{\mu\nu\rho\sigma}+(boundary\ term)\\
&\label{a3}\epsilon^{\mu\nu\rho\sigma} \epsilon^{IJKL} e_{\mu I}e_{\nu J}R_{\rho\sigma KL}=(d^4x)e[R(e)+K_{\mu\nu\rho}K^{\mu\nu\rho}-K^\mu\,_{\mu\rho}K_\nu\,^{\nu\rho}]+(boundary\ term)\,.
\end{align}
%\section*{References}


\begin{thebibliography}{0}
%\cite{Hojman:1980kv}
\bibitem{Hojman:1980kv} 
  R.~Hojman, C.~Mukku and W.~A.~Sayed,
  ``Parity Violation In Metric Torsion Theories Of Gravitation,''
  Phys.\ Rev.\ D {\bf 22}, 1915 (1980).
  %%CITATION = PHRVA,D22,1915;%%

%\cite{Perez:2005pm}
\bibitem{Perez:2005pm} 
  A.~Perez and C.~Rovelli,
  %``Physical effects of the Immirzi parameter,''
  Phys.\ Rev.\ D {\bf 73}, 044013 (2006)
  [gr-qc/0505081].
  %%CITATION = GR-QC/0505081;%%

%\cite{Freidel:2005sn}
\bibitem{Freidel:2005sn} 
  L.~Freidel, D.~Minic and T.~Takeuchi,
  %``Quantum gravity, torsion, parity violation and all that,''
  Phys.\ Rev.\ D {\bf 72}, 104002 (2005)
  [hep-th/0507253].
  %%CITATION = HEP-TH/0507253;%%

%\cite{Mercuri:2006um}
\bibitem{Mercuri:2006um} 
  S.~Mercuri,
  %``Fermions in Ashtekar-Barbero connections formalism for arbitrary values of the Immirzi parameter,''
  Phys.\ Rev.\ D {\bf 73}, 084016 (2006)
  [gr-qc/0601013].
  %%CITATION = GR-QC/0601013;%%

%\cite{Kaul:2007gz}
\bibitem{Kaul:2007gz} 
  R.~K.~Kaul,
  ``Holst Actions for Supergravity Theories,''
  Phys.\ Rev.\ D {\bf 77}, 045030 (2008)
  [arXiv:0711.4674 [gr-qc]].
  %%CITATION = ARXIV:0711.4674;%%

%\cite{Durka:2009pf}
\bibitem{Durka:2009pf} 
  R.~Durka, J.~Kowalski-Glikman and M.~Szczachor,
  ``Supergravity as a constrained BF theory,''
  Phys.\ Rev.\ D {\bf 81}, 045022 (2010)
  [arXiv:0912.1095 [hep-th]].
  %%CITATION = ARXIV:0912.1095;%%

%\cite{Barbero:1994ap}
\bibitem{Barbero:1994ap} 
  J.~F.~Barbero G.,
  ``Real Ashtekar variables for Lorentzian signature space times,''
  Phys.\ Rev.\ D {\bf 51}, 5507 (1995)
  [gr-qc/9410014].
  %%CITATION = GR-QC/9410014;%%

%\cite{Immirzi:1996di}
\bibitem{Immirzi:1996di} 
  G.~Immirzi,
  ``Real and complex connections for canonical gravity,''
  Class.\ Quant.\ Grav.\  {\bf 14}, L177 (1997)
  [gr-qc/9612030].
  %%CITATION = GR-QC/9612030;%%

%\cite{Ashtekar:1986yd}
\bibitem{Ashtekar:1986yd}
  A.~Ashtekar,
  ``New Variables for Classical and Quantum Gravity,''
  Phys.\ Rev.\ Lett.\  {\bf 57} (1986) 2244.
  %%CITATION = PRLTA,57,2244;%%

%\cite{Wise:2009fu}
\bibitem{Wise:2009fu} 
  D.~K.~Wise,
  ``Symmetric space Cartan connections and gravity in three and four dimensions,''
  arXiv:0904.1738 [math.DG].
  %%CITATION = ARXIV:0904.1738;%%


%\cite{Holst:1995pc}
\bibitem{Holst:1995pc} 
  S.~Holst,
  ``Barbero's Hamiltonian derived from a generalized Hilbert-Palatini action,''
  Phys.\ Rev.\ D {\bf 53}, 5966 (1996)
  [gr-qc/9511026].
  %%CITATION = GR-QC/9511026;%%

%\cite{Benedetti:2011nd}
\bibitem{Benedetti:2011nd} 
  D.~Benedetti and S.~Speziale,
  ``Perturbative quantum gravity with the Immirzi parameter,''
  JHEP {\bf 1106}, 107 (2011)
  [arXiv:1104.4028 [hep-th]].
  %%CITATION = ARXIV:1104.4028;%%

%\cite{Neville:1979rb}
\bibitem{Neville:1979rb} 
  D.~E.~Neville,
  ``Gravity Theories With Propagating Torsion,''
  Phys.\ Rev.\ D {\bf 21}, 867 (1980).
  %%CITATION = PHRVA,D21,867;%%

%\cite{de Wit:1982ig}
\bibitem{de Wit:1982ig} 
  B.~de Wit and H.~Nicolai,
  ``N=8 Supergravity,''
  Nucl.\ Phys.\ B {\bf 208}, 323 (1982).
  %%CITATION = NUPHA,B208,323;%%

%\cite{Ferrara:1976kg}
\bibitem{Ferrara:1976kg} 
  S.~Ferrara, D.~Z.~Freedman, P.~van Nieuwenhuizen, P.~Breitenlohner, F.~Gliozzi and J.~Scherk,
  ``Scalar Multiplet Coupled to Supergravity,''
  Phys.\ Rev.\ D {\bf 15}, 1013 (1977).
  %%CITATION = PHRVA,D15,1013;%%

%\cite{VanNieuwenhuizen:1979hma}
\bibitem{VanNieuwenhuizen:1979hma} 
  P.~van Nieuwenhuizen,
  ``Supergravity As A Gauge Theory Derived From Matter Coupling,''
  ITP-SB-79-93.
  %%CITATION = ITP-SB-79-93;%%

%\cite{Ferrara:1976ni}
\bibitem{Ferrara:1976ni} 
  S.~Ferrara, F.~Gliozzi, J.~Scherk and P.~Van Nieuwenhuizen,
  ``Matter Couplings in Supergravity Theory,''
  Nucl.\ Phys.\ B {\bf 117}, 333 (1976).
  %%CITATION = NUPHA,B117,333;%%




\end{thebibliography}
\end{document}